\documentclass[conference]{IEEEtran}
\IEEEoverridecommandlockouts
\pdfoutput=1

\usepackage{cite}
\usepackage{amsmath,amssymb,amsfonts}
\usepackage{algorithmic}
\usepackage{graphicx}
\usepackage{textcomp}
\usepackage{xcolor}
\usepackage{multirow} 
\usepackage{siunitx} 
\usepackage{listings} 
\usepackage{subcaption} 
\usepackage{pgfplots} 
\usepackage{optidef} 
\usepackage[inline,shortlabels]{enumitem} 
\usepackage[noabbrev]{cleveref}
\definecolor{mycolor1}{rgb}{0.00000,0.44700,0.74100}%
\definecolor{mycolor2}{rgb}{0.85000,0.32500,0.09800}%
\definecolor{mycolor3}{rgb}{0.92900,0.69400,0.12500}%
\definecolor{mycolor4}{rgb}{0.49400,0.18400,0.55600}%
\def\BibTeX{{\rm B\kern-.05em{\sc i\kern-.025em b}\kern-.08em
    T\kern-.1667em\lower.7ex\hbox{E}\kern-.125emX}}

\usepackage{flushend}
\usepackage{epstopdf}

\begin{document}
\title{Design and validation of a low cost programmable growth chamber for study and production of plants, mushroom, and insect larvae\\
\thanks{This project has received funding from the European Social Fund (ESF)}
}

\author{\IEEEauthorblockN{Murali Padmanabha}
\IEEEauthorblockA{\textit{Technische Universitaet Chemnitz}\\
	\textit{Automatic Control and System Dynamics Lab} \\
Chemnitz, Germany \\
murali.padmanabha@etit.tu-chemnitz.de}
\and
\IEEEauthorblockN{Stefan Streif}
\IEEEauthorblockA{\textit{Technische Universitaet Chemnitz}\\
	\textit{Automatic Control and System Dynamics Lab} \\
	Chemnitz, Germany \\
	stefan.streif@etit.tu-chemnitz.de}
}
\newcommand{\citep}{\cite}
\maketitle
\graphicspath{{gfx/pdf/}}
\begin{abstract}
Use of commercial growth chambers for study of biological processes involved in biomass growth and production pose certain limitations on the nature of studies that can be performed in them. Optimization of biomass rearing and production process requires quantitative study of environment influences on the organism and eventually the products and byproducts consumed and produced.
 This work presents a low cost modular system designed to facilitate quantitative study of growth processes and resource exchanges in organisms such as plants, fungi and insect larvae.
 The proposed system constitutes of modular units each performing a specific function. A novel compact thermoelectric cooler based unit is designed for conditioning the air. Sensor cluster for measuring gas concentrations, air properties (temperature, humidity, pressure), and growing medium properties is implemented and tested. An actuator cluster for resource exchange and a wiring and control scheme for light spectrum adjustment is proposed. A three tier hierarchical software framework consisting of an open-source cloud platform for data aggregation and user interaction, embedded firmware for microcontroller, and an application development framework for test automation and experiment regime design is developed and presented.   
 A series of experiments and tests were performed using the designed hardware and software to evaluate its capabilities and limitations. This controlled environment was used to study the photosynthesis and its dependency on temperature and light intensity in \textit{Ocimum basilicum}. In a second experiment, evolution of metabolic activity of \textit{Hermetia illucens} larvae over its larval phase was studied and the metabolic products and byproducts were quantitatively measured.
\end{abstract}

\section{Introduction}
The problems associated with food production due to the growing population, changing climate, reduced ground water resources, increased transportation costs could be solved using concepts such as vertical farming \citep{DESPOMMIER2011}, urban agriculture, and plant factories with artificial lights (PFAL) \citep{KOZAI2013}. These modern food production techniques proves to be effective in increasing the biomass throughput per volume of water used per growing area in comparison to the conventional farming techniques. These methods however, require significant amount of energy for generating the artificial micro-climate necessary for the plants growing in them \citep{ALCHALABI201574}.

Several studies have been undertaken to evaluate the economic feasibility of a typical vertical farm and how these farms could be made profitable by combining different organisms (e.g. plant and fishes) and exploiting the symbiotic behaviour between them. Such investigations are noticeable in the area of space research and exploration projects for designing bio-regenerative life-support systems: MELiSSA \citep{LASSEUR2005}, ACLS \citep{BOCKSTAHLER2017} CELSS \citep{CARY1994}, and CAB \citep{LOBASCIO2008}. Improved biomass output was shown when plants and mushrooms are grown in symbiosis \citep{KITAYA1994}. Studies performed on farms with plant-fish integrated production have also shown reduced operation cost \citep{Jagath2010}. The economic feasibility analysis of the vertical farm performed by the German Aerospace Center DLR has shown that on combining production of different organisms reduces the overall cost in such farms \citep{CONRAD2017}.
Quantification of mass and resource fluxes between organisms is important for performing simulation studies, designing experiments and developing automation and farm infrastructure. CUBES Circle is another project, funded by the German Federal Ministry of Education and Research, aiming on the study of mass and energy fluxes between plants, fishes and insects connected together \citep{CUBESCircle2018}.

The above discussed works highlight some of the research potential and shortfalls in the area of optimized food production in controlled and connected environments . This requires firstly, a quantitative study of growth processes of individual organisms and its environmental factors. Secondly, study of resource fluxes and symbiosis between organisms and its environment. Finally, development of mathematical models that adequately describe them.

The fundamental step of such studies is the development of a testbench suitable for different organisms. Plants (\textit{Lactuca sativa}, \textit{Ocimum basilicum}), fungi (\textit{Agaricus bisporus}, \textit{Lentinula edodes}, etc.), and insect larvae (\textit{Hermetia illucens}, \textit{Tenebrio molitor}) are some of the candidates considered in this work. A preliminary analysis of the requirements for developing such system highlights the need for a sophisticated setup with actuators to generate the necessary micro-climate, sensors to measure various variables (e.g. O$_2$, CO$_2$, humidity, etc.), and interface to exchange resources (e.g. water, air, nutrients). Some modern commercial growth chambers support CO$_2$ concentration regulation and inlet for additional sensors through instrument port but infrastructure for resource exchange and test automation is either limited or non existent \citep{CONVIRON2019}. Modification of these growth chambers for incorporating necessary infrastructure is theoretically possible but adds to the cost and complexity. Literatures \citep{PFC2019,ZABEL20161} also indicate the need for custom devices irrespective of the availability of commercial growth chambers for studying and phenotyping biological organisms. This is primarily due to a very high equipment cost ($>6000$ EUR for growth chamber and additional cost for control software), proprietary hardware/software, and requirements very specific to the nature of studies performed.

This work uses some of the concepts proposed in the literature and proposes novel ideas and modifications to develop a low cost system suitable for previously mentioned studies.
The following are some of the design requirements considered and realized:
\begin{itemize}
		\item Actuators for air conditioning, light spectrum adjustment, and on-demand water and gas exchange with airtight containment.
		\item Measurement system for air quality, gas concentrations, air pressure, air and substrate temperatures, humidity and substrate moisture.
		\item Modular design using off-the-shelf components and 3D printable parts.
		\item Open-source software framework for experiment regime design and execution, user interaction, data collection and analysis.
\end{itemize}


In the following sections, firstly hardware component selection, electronic interface design are explained followed by software component design and integration of an example control regime. Results of sensor-actuator tests for performance evaluation of the designed system is presented. Finally, results of experiments executed for the study of growth of biological organisms are discussed.


\section{Material and Methods}
The rearing process of biological organism such as plants, fungi and insects have special requirements with respect to climate, nutrient supply and gas concentrations. This section describes in detail how these special requirements were translated into design choices using off-the-shelf hardware components, 3D-Printable custom parts and open-source software.

\subsection{Hardware Design}\label{sec:design_chamber}
\begin{figure*}[h]
	\centering
	\resizebox{0.95\textwidth}{!}{\includegraphics{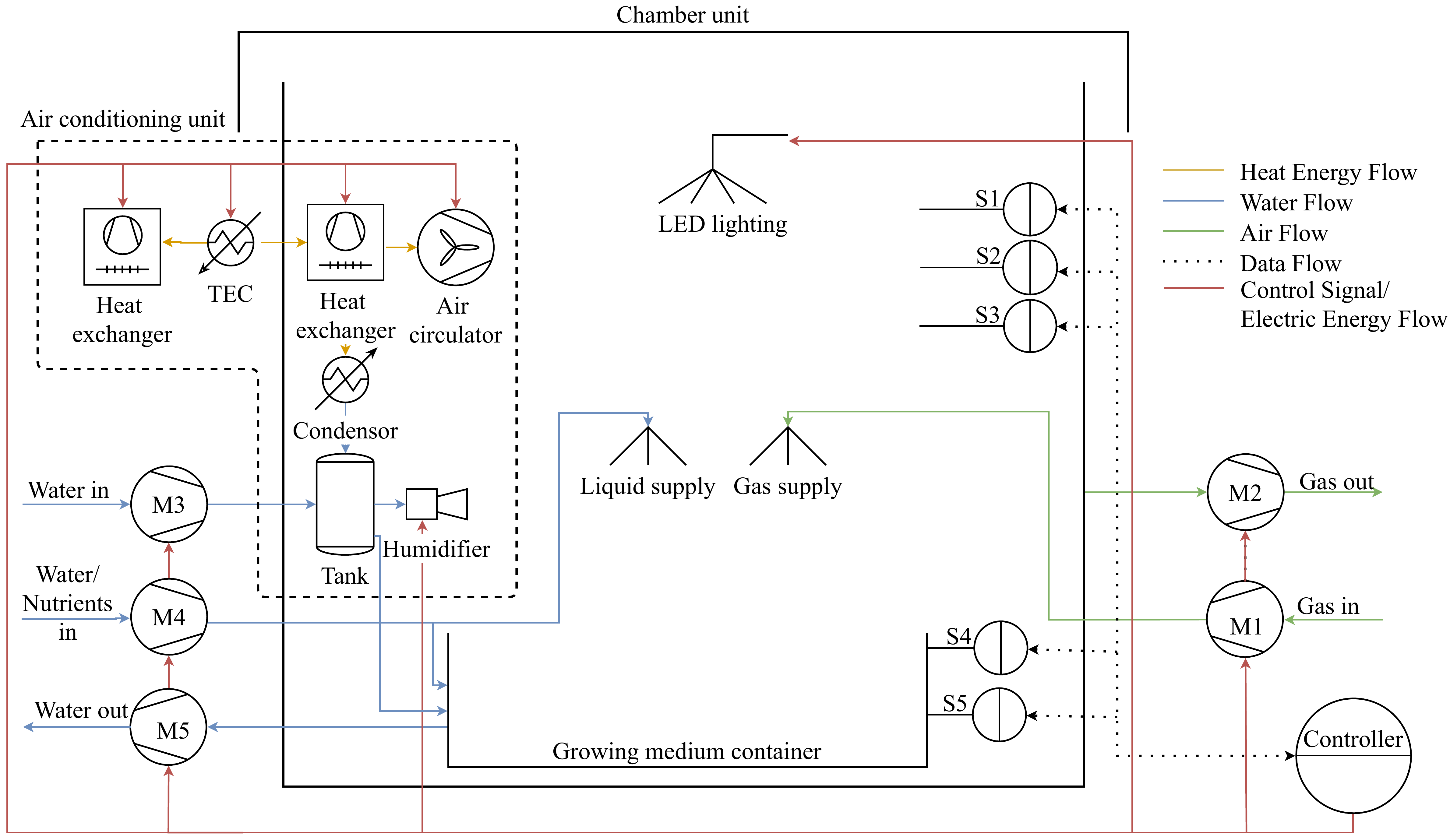}}
	\caption[Plant Model]{Process model of the controlled environment consisting of central chamber unit, air conditioning unit, liquid and gas circulation pumps (M1-M5), LED light panel, sensors (S1-S5) and central embedded controller.}
	\label{fig:plant_model}
\end{figure*}

A process model of a growth chamber, as shown in Fig.~\ref{fig:plant_model}, was drafted based on the functional requirements. The central chamber unit is where the organism of interest is reared and studied. This part being the fundamental component of this process model, was selected first. A polypropylene container of \SI{75}{\liter} volume with airtight sealable lid was selected. This volume was constrained due to the choice of an off-the-shelf container which also corresponds to the volume of low-mid tier commercial growth chamber. Based on this chosen volume, specification of other modules (e.g. heater, cooler, pumps) were derived. Modifications required by the chamber to obtain desired functionalities are explained in the following sections.



\subsubsection{Air conditioning}\label{heating_cooling}
Temperature of the growing chamber in which subjects grow, influences various metabolic processes and also triggers certain biological events. In case of \textit{Lentinula edodes} fungi, fruiting is triggered by a temperature drop \citep{Philippoussis2003}. Similarly in case of plants, biological processes such as, photosynthesis and even onset of anthesis is affected by temperature \cite{John1990}. Humidity also influences the metabolic activities of both the subject and unwanted fungus. Humid environment with sufficiently warm temperature provides suitable platform for organisms such as unwanted fungus to thrive and possibly suppress or destroy the growth of the subject \citep{BLOCK1953}. Therefore it is important to condition the air inside the chamber. 

Application of compact ceramic heating elements for heating air can be seen widely in industrial and home appliances\citep{KULWICKI1984,Paganelli1984}. However, cooling the air requires sophisticated mechanical parts and increases the size and cost. Research works \citep{CHEIN20042207,RIFFAT20041979} showcase the application of thermoelectric cooler (TEC) module for heating, cooling and de-humidification.

Humidification on the contrary can be achieved by accelerating the evaporation of water. Despite the transpiration contributed by metabolic activities of the subject growing in the chamber, it requires long duration to reach desired humidity and is uncontrolled. To compensate for this shortcoming, the humidification process can be controllably accelerated using a ultrasonic atomizer that breaks water into fine particles \citep{TOPP1972127}. These technologies were combined together to implement the air conditioning system.

\paragraph{Construction}
A combination of TEC module of varying cooling capacities (\SI{36}{\watt}, \SI{75}{\watt}, \SI{126}{\watt}) were tested with heat exchanging components of different sizes for both hot and cold side. A \SI{126}{\watt} module with peak current and voltage of \SI{14.7}{\ampere} and \SI{14.5}{\volt} respectively, was selected for highest performance. The side of the TEC module facing the inner side of the chamber is interfaced with an aluminium passive heat exchanger while the opposite side with an active water cooled heat exchanger.
A combination of two PC-cooling fans are used in push-pull arrangement to actively cool the otherwise passive heat exchanger and also to introduce air circulation inside the chamber unit. This design choice of non-similar heat exchanger is made to improve the cooling/heating performance while maximizing the inner volume of the chamber unit.
In cooling mode of operation, the inner heat exchanger condenses water which is accumulated in a reservoir. An ultrasonic atomizer is integrated into the reservoir  using the water stored in it for humidification. The enclosure with integrated reservoir that encapsulates the passive heat exchanger, air circulation fans, and also the humidifier is custom designed and 3D printed. These parts as depicted within dashed box in Fig.~\ref{fig:plant_model} constitutes the air conditioning unit.


\paragraph{TEC Driver}
\begin{figure}[!h]
	\centering
	\resizebox{0.485\textwidth}{!}{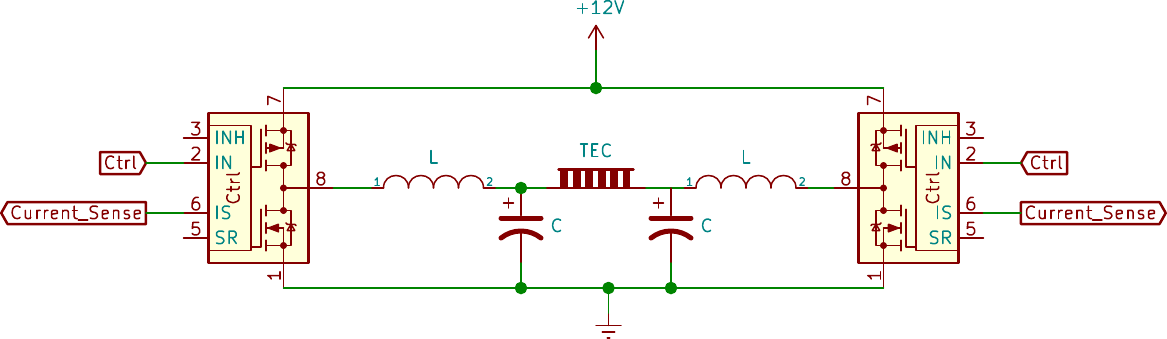}
	\caption[System Arch]{Electronic driver for TEC comprising a H-bridge constructed from dual BTN8982 half bridge and low-pass L-C filter for digital to analog conversion}\label{fig:tec_sch}
\end{figure}
Operation of a TEC module for both heating and cooling requires the infrastructure for changing the magnitude and polarity of the voltage applied across it. Change of voltage polarity can be achieved using a H-bridge circuit but magnitude variation requires additional circuitry. This driver circuit, as seen in Fig.~\ref{fig:tec_sch}, can be constructed using two half bridge BTN8982 device and a low-pass L-C filter\citep{Andria2016}. This combination enables the change of current flow direction through the TEC thus changing its operation mode. Using PWM signals to operate the driver in ON-OFF mode, enables the control of current flowing through the device. However, the power loss in TEC module in such a configuration is higher than in DC mode. To compensate this power loss, a low-pass L-C filter circuit with a cut-off frequency $f_c$ equal to  \SI{11.813}{KHz} is implemented between the TEC and the H-bridge driver. A source PWM frequency of \SI{64}{KHz} is generated from microcontroller such that the voltage ripples are eliminated from the filter circuit resulting in a smooth DC voltage.

\subsubsection{Gases, water and nutrient exchange}
The primary raw and by-products of the biological processes constitute O$_2$, CO$_2$ and H$_2$O (liquid and vapour). Concentration measurement of these compounds in gaseous forms is necessary for quantitative study of the mass fluxes due to the underlying biological processes. This requires measurements to be performed in sealed chamber. Simultaneously to enable normal growth, these concentrations must be regulated. Adding to the complexity, water and nutrient supply mechanism for the subjects vary with some requiring overhead spray (fungi) and some through a water bath. These requirements pose a challenge to design the infrastructure that provides both air-tight containment and fluid exchange on demand.
To overcome this limitation, a circulation unit constituting a total of 5 membrane pumps M1-M5 are used for fluid circulation to and out of the chamber. These pumps enable the circulation on demand while keeping the chamber air-tight.
Pumps M1 and M2 are coupled to work in opposite flow directions such that M1 removes the gas mixture from the chamber to the external sink and M2 fills it with gases of known concentrations from an external source. For water and nutrient supply the circulation unit uses pumps M3-M5, where M3 supplies water to internal reservoir and eventually to growing medium container on overflow, M4 supplies water and nutrient mixture to spray nozzle or directly to growing medium container, and M5 removes water to external reservoir. Thus, completing the water and gas circulation between chamber and the external source.
\subsubsection{Lighting}
\begin{figure*}[!h]
	\centering
	\begin{subfigure}[c]{0.49\textwidth}
		\centering
		\resizebox{\columnwidth}{!}{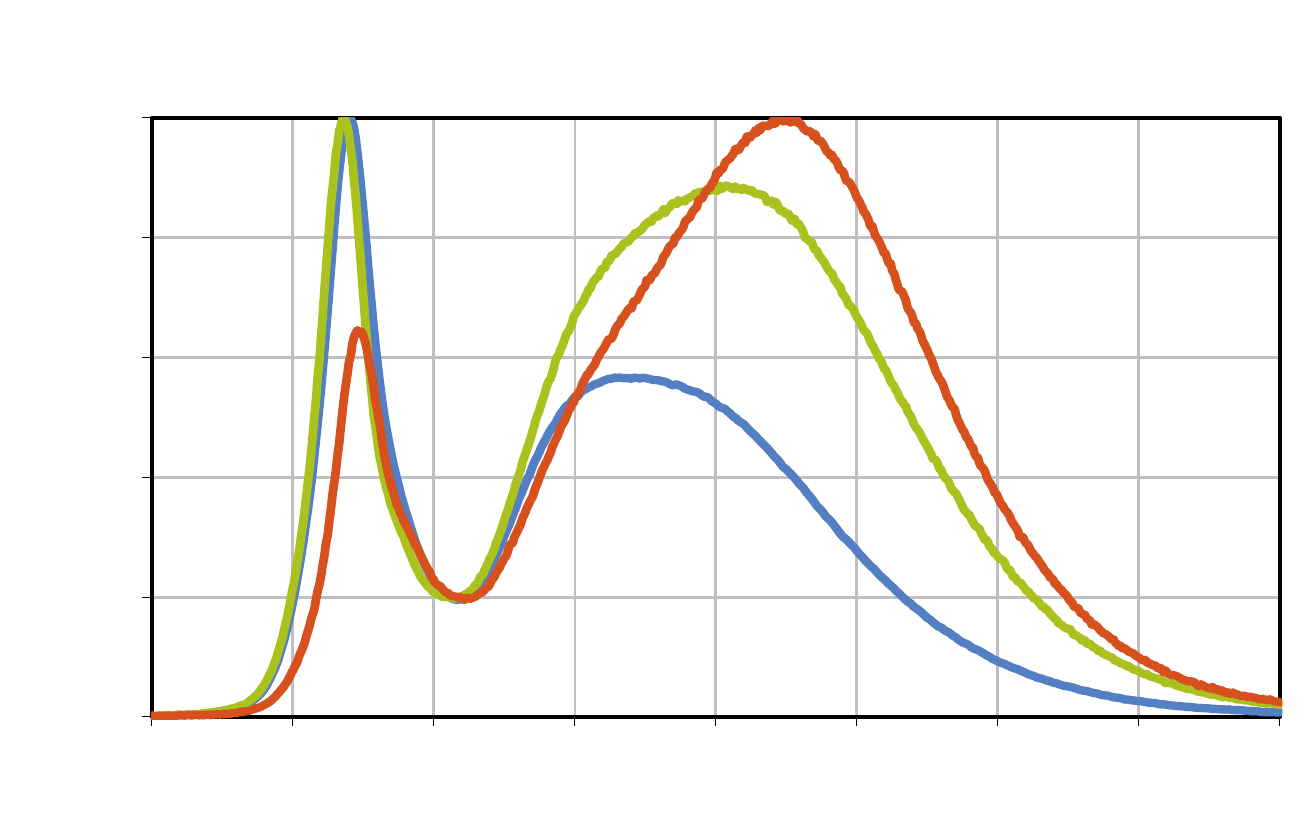}
	\end{subfigure}
	\begin{subfigure}[c]{0.49\textwidth}
		\centering
		\resizebox{\columnwidth}{!}{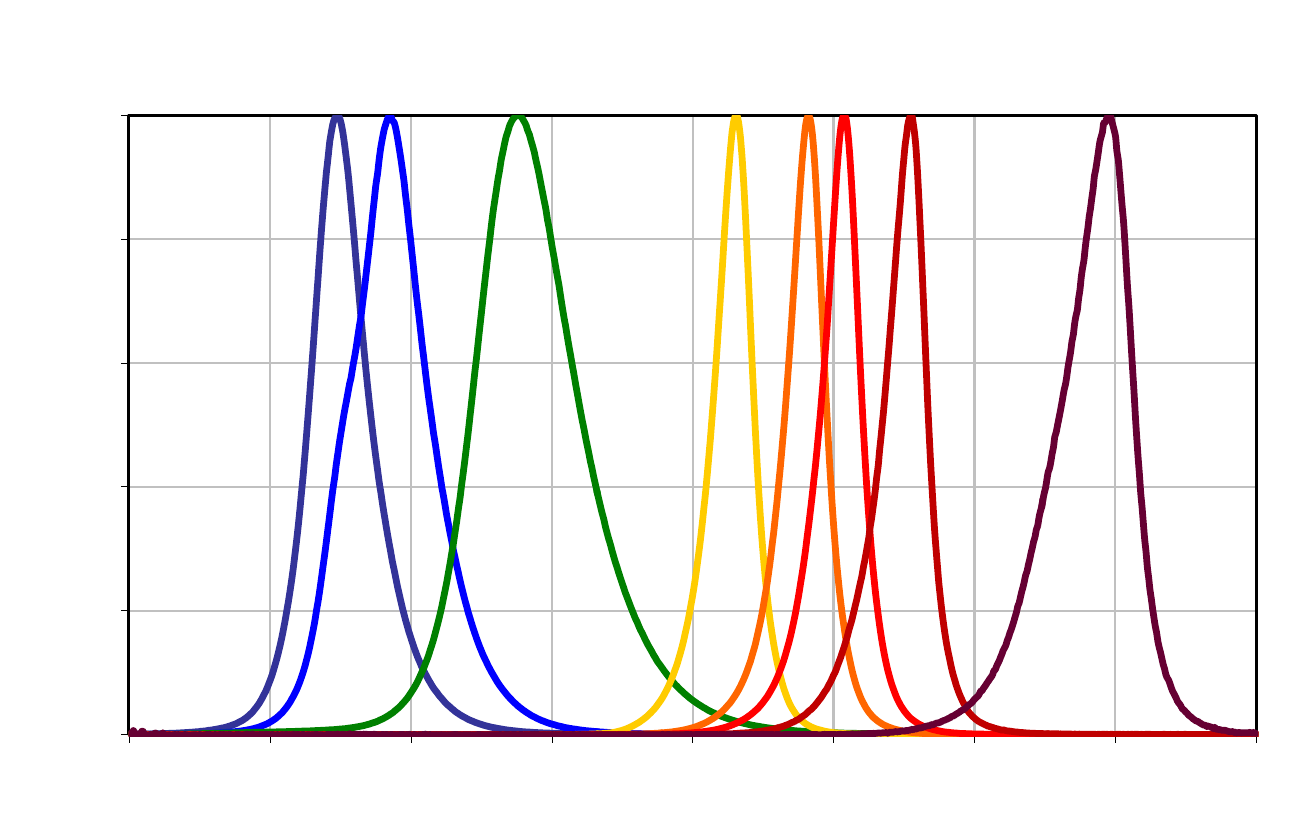}
	\end{subfigure}
	\caption{Normalized spectral power distribution of LEDs available for lighting system design. White LED series compared on left \citep{XPELED2018} and narrow-band LED series compared on right \citep{MCELED2018}.}
	\label{fig:spectrum_LED}
\end{figure*}

The radiant flux required for the subjects of interest varies between species and organisms. Certain fungal species require white light while plant species mostly require red and blue components\citep{OGUNTOYINBO2013340}. Literature have also indicated the changes in chemical composition in plants in response to red and far-red light\citep{Schwend2016243,KIM2004143}. Events such as flowering and germination could be triggered by varying the spectral components and circadian rhythms\citep{Schwend2016148,Massa01122008}.  This, therefore requires light source with adjustable spectral composition and power. Narrow-band LEDs covering a wide range of wavelengths and wide-band white LEDs as seen in Fig.~\ref{fig:spectrum_LED} are widely available from different manufacturers\citep{XPELED2018,MCELED2018}.

One of the goals of this work was to design and include a light source with adjustable spectrum. Spectral power distribution of typical white LEDs with different colour temperature (3000-6500 \si{\kelvin}) could be used as the primary source and the additional wavelengths that influences various biological processes or trigger events could be included using narrow-band LEDs. A complete range of spectral requirements for various species is still unknown and needs to be studied. A preliminary LED lighting unit with a combination of neutral white, narrow-band red, blue, green, and support for inclusion of UV/far-blue and IR/far-red LEDs was designed. 

\begin{figure}[!h]
	\centering
	\resizebox{0.485\textwidth}{!}{\includegraphics{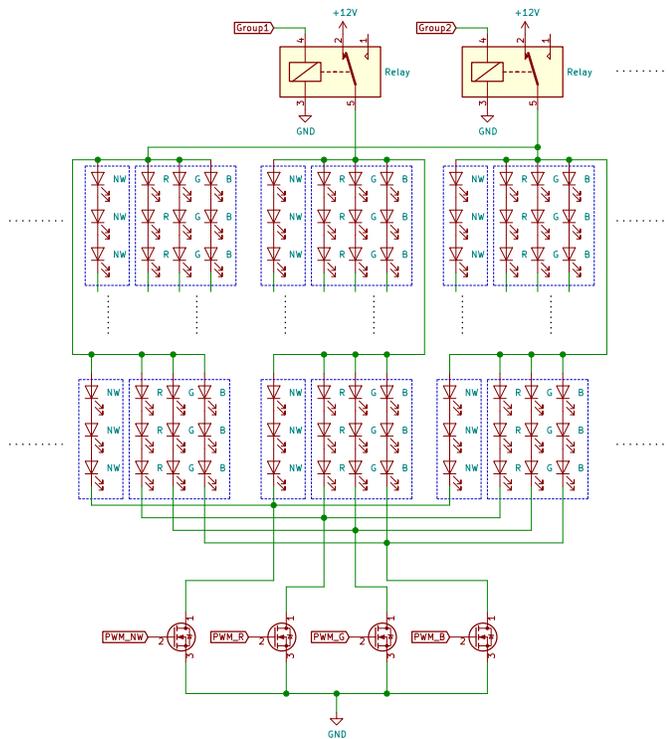}}
	\caption[LED Panel Construction]{LED panel wiring schematic: The LEDs grouped in the blue dashed boxes indicate the 12V modules. Horizontal dotted line indicate the LED groups and the vertical dotted line indicates the daisy-chaining of LED channel modules.}
	\label{fig:led_sch}
\end{figure}

\paragraph{LED Driver for Spectrum regulation}
Spectrum adjustment and light-based event triggering require that the individual channels (i.e. white, red, blue, green, UV/far-blue and IR/far-red) be independently adjusted. LEDs used in this system are daisy-chained modules of 3 or 4 LEDs connected in series and operating at \SI{12}{\volt}. As seen in Fig.~\ref{fig:led_sch}, every colour channel has a MOSFET based low-side driver to adjust the radiant spectral power using PWM signals. These vertical columns of daisy-chained modules are combined into groups that are operated using relays. This grouping is done such that the first group corresponds to the central vertical column and the preceding groups correspond to the adjacent columns.
The overall advantage of this setup is that the radiant spectral power and the total area covered by the LEDs can be varied as required.

\subsubsection{Sensors}
A combination of sensors S1-S5, listed in Table \ref{tab:sensor_spec}, are capable of measuring CO$_2$, O$_2$, and volatile organic compound (VOC) concentrations, air and substrate temperatures, atmospheric pressure, relative humidity, and moisture of growing medium. Sensor S1 is a non-dispersive infrared sensor for CO$_2$ concentration measurement also housing an additional relative humidity and temperature sensing element. Sensor S2 is an environmental sensor for measuring relative humidity, temperature, pressure, and VOC concentration. The O$_2$ sensor S3 is a electro-chemical galvanic cell that produce voltage on exposure to O$_2$. Its sensitivity is proportional to the O$_2$ concentration but its analogue output is out of range of the internal ADC of the microcontroller and thus an external ADC with programmable gain is used for signal amplification. All the above mentioned sensors use I2C communication interface to transfer measurement data to the microcontroller and is enclosed as a stand-alone sensor unit.
\begin{table}[!h]
	\centering
	\caption{List of sensors and sensor specification}
	\resizebox{0.485\textwidth}{!}{
		\begin{tabular}{lllllllll}\hline\\
			ID & Part No. & Manufacturer & Interface & Parameters & Min   & Max   & Accuracy & Unit \\ \hline\\
			\multirow{3}[0]{*}{S1} & \multirow{3}[0]{*}{SCD30} & \multirow{3}[0]{*}{Sensirion} & \multirow{3}[0]{*}{I2C} & CO$_2$   & 400   & 10000 & 30    & PPM \\
			&   &  &  & Temperature & -40   & 70    & 0.3   & \SI{}{\celsius} \\
			&   &  &  & Humidity & 0     & 100   & 0.1   & \%RH \\\hline \\
			\multirow{3}[0]{*}{S2} & \multirow{3}[0]{*}{BME680} & \multirow{3}[0]{*}{BOSCH} & \multirow{3}[0]{*}{I2C} & Temperature & -40   & 85    & 1     & \SI{}{\celsius} \\
			&   &  &  & Humidity & 10    & 90    & 3     & \%RH \\
			&   &  &  & Pressure & 300   & 1100  & 0.6   & hPa \\\hline \\
			S3    & SK25-F & Figaro & Analog & O$_2$    & 0     & 30    & 1     & \% \\\hline \\
			S4    & DS18B20 &Dallas Semiconductor& 1-Wire& Temperature & -55   & 125   & 0.5   & \SI{}{\celsius} \\\hline \\
			S5    & SEN0193 & DFRobot & Analog& Moisture & 1.2   & 2.5   & -     & V \\ \hline
		\end{tabular}%
		}\label{tab:sensor_spec}%
\end{table}
Temperature sensor S4 is used for substrate temperature measurement and communicates with the controller using a proprietary one-wire interface. This interface allows addition of several sensors to the same data-bus without additional hardware changes. The substrate moisture concentration is measured using a capacitive sensor S5 with analog voltage output. These two sensors are separated from the other sensors to facilitate their placement inside the substrate or the growing medium.

\subsubsection{Control Unit}
\begin{figure}[h]
	\centering
	\resizebox{0.485\textwidth}{!}{\includegraphics{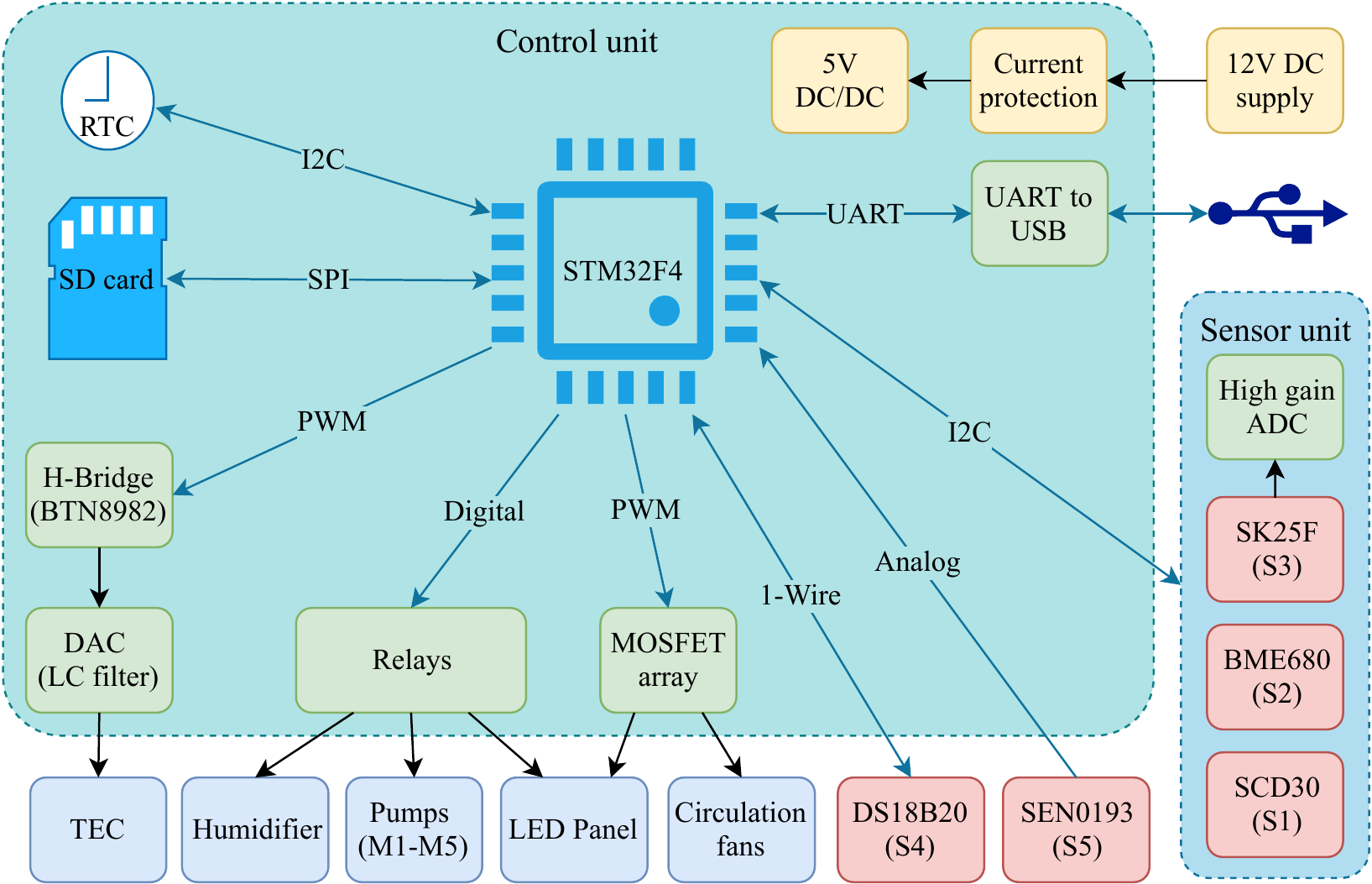}}
	\caption[LED Panel Construction]{Electronic components and communication interfaces: internal components of the control and sensor unit, external components and the corresponding communication and electrical interfaces.}
	\label{fig:ele_layout}
\end{figure}

The control unit, as seen in Fig.~\ref{fig:ele_layout}, constitutes the microcontroller, drivers, switching circuits, voltage regulation and current protection circuits. The microcontroller used is a 32-bit ARM Cortex M4 device from ST-Microelectronics clocked at \SI{100}{\mega\hertz}. It is connected to a SD card for local data logging and storing configurations and calibration data. A real-time clock (RTC) is used for time keeping and timestamps for logged data. Communication to external system is realized using UART to USB interface. The PWM peripheral pins drives the LED lighting system, TEC heating-cooling system, internal circulation fans and external heat exchanger fan. General purpose input-output drives the electromechanical relays of the led group switch and pumps M1-M5. The entire system is designed to be operated using a single \SI{12}{\volt} source and thus incorporates the necessary DC-DC step-down converter for supplying \SI{5}{\volt} for low voltage electronic components (microcontroller, humidifier, etc.).

\subsubsection{Assembled System}
\begin{figure*}[!h]
	\centering
	\resizebox{\textwidth}{!}{\includegraphics{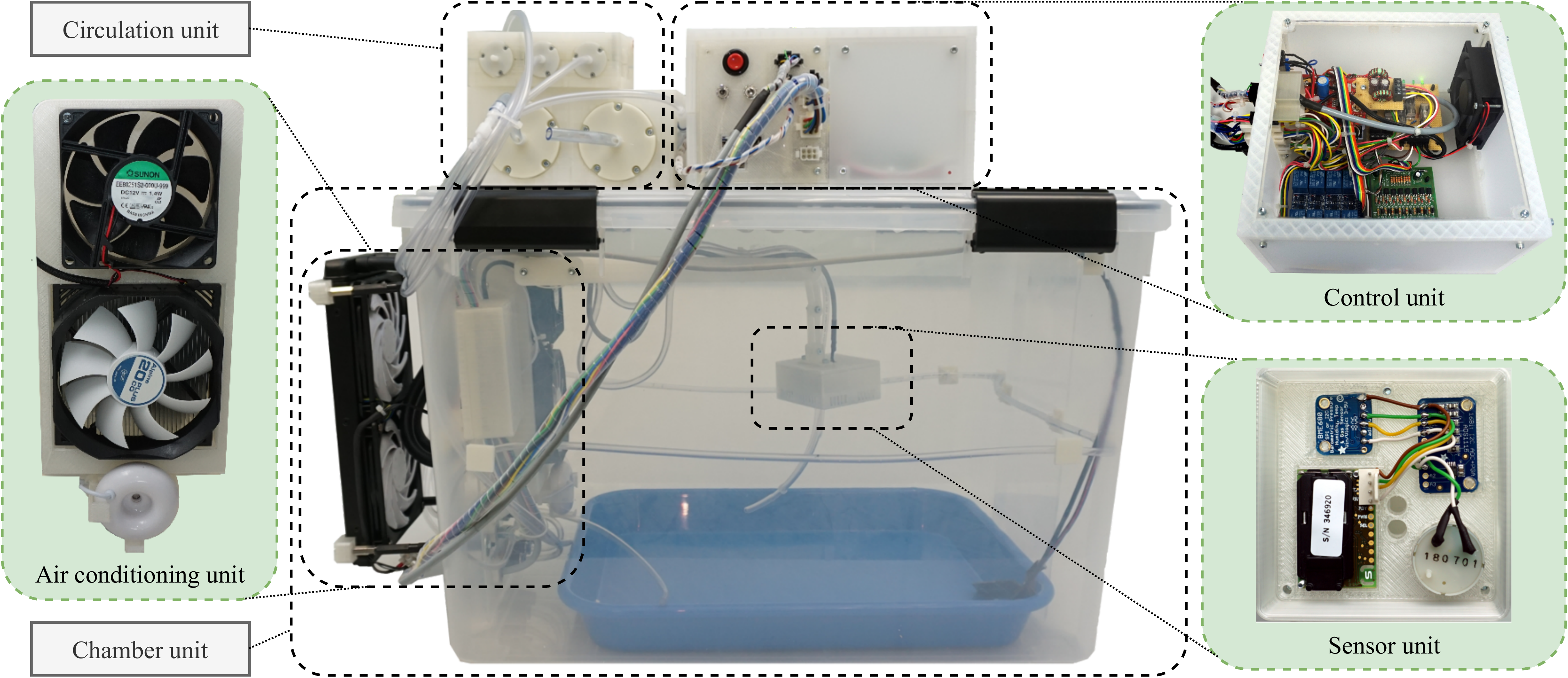}}
	\caption[Plant Model]{Growth chamber with the assembled units and their internal view (highlighted in green).}
	\label{fig:test_bench}
\end{figure*}

Components constituting the modular units were assembled together with 3D printed parts and enclosures as seen in Fig.~\ref{fig:test_bench}. The control unit enclosure is made of a 3D printed frame and acrylic walls with Molex connectors for connecting it with the chamber unit. The sensor unit is suspended inside the chamber unit with a 3D printed adjustable arm. The circulation unit is connected with the bulkhead connectors of the chamber unit using a 4mm polyvinyl chloride tubes, keeping the chamber unit air tight. LED panel, not visible in the figure, is mounted on the inner side of the chamber lid and connected using a Molex connector to enable complete disconnection while opening the lid. Push-pull arrangement of the fans and the integrated humidifier can be seen in the highlighted air conditioning unit in Fig.~\ref{fig:test_bench}.

\subsection{Software Design}
Similar to the modular approach of the hardware design, software components were also developed as independent components and integrated using standard interfaces. To enable autonomous operation of the chamber and machine-to-machine (M2M) communication, the following requirements were established and realized:
\begin{itemize}
	\item Design and execution of control application or custom experiment regimes on the designed system.
	\item Access to measurement data and system states for performing data analytics, monitoring and process control.
	\item Expert knowledge input for optimal chamber operation.
	
\end{itemize}

\begin{figure*}[h]
	\centering
	\resizebox{\textwidth}{!}{\includegraphics{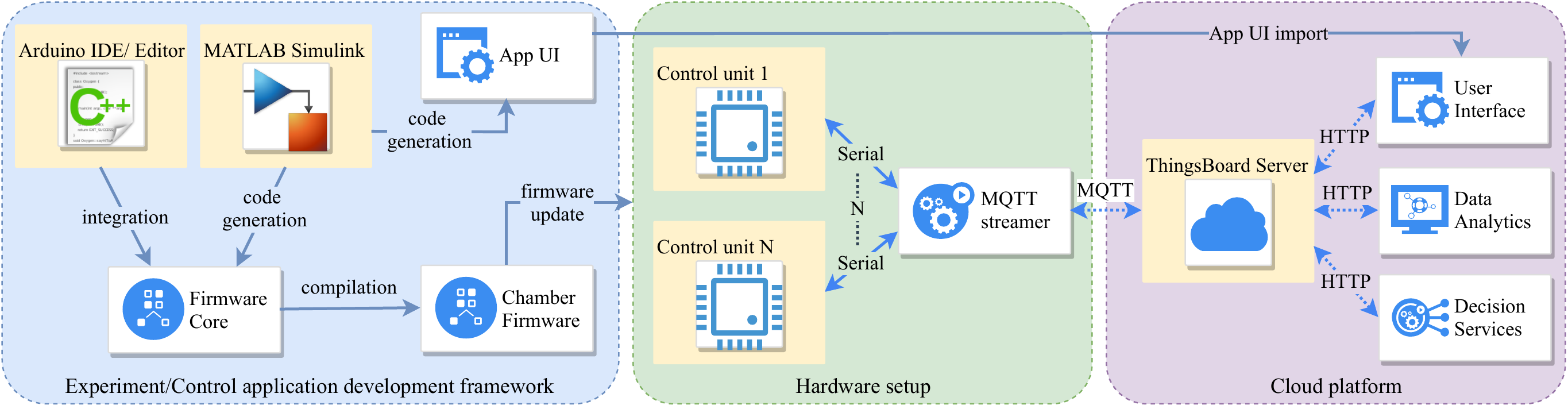}}
	\caption[System Arch]{Software and information exchange framework: consisting of code generation framework for application development, hardware setup for experiment and data transmission, and cloud platform for data collection and user interaction.}
	\label{fig:sys_arch}
\end{figure*}

These requirements were translated to a framework, as seen in Fig.~\ref{fig:sys_arch}, with a three part solution. \begin {enumerate*} [1) ]%
\item Open-source firmware core for control unit with necessary drivers, middleware and application interface for reliable operation of the chamber.
\item Cloud platform with data, protocol and web sever for data aggregation, data visualization and interaction with chambers for process control.
\item Code generation framework for control application or experiment regime development. \end {enumerate*}
This architecture facilitates the use of the controlled evironment for different subjects and executing different experiments. 
The following sections explain the software design of the IoT Framework and device firmware in detail.

\subsubsection{Device Firmware}\label{sec:design_firmware}
\begin{figure}[h]
	\centering
	\resizebox{0.485\textwidth}{!}{\includegraphics{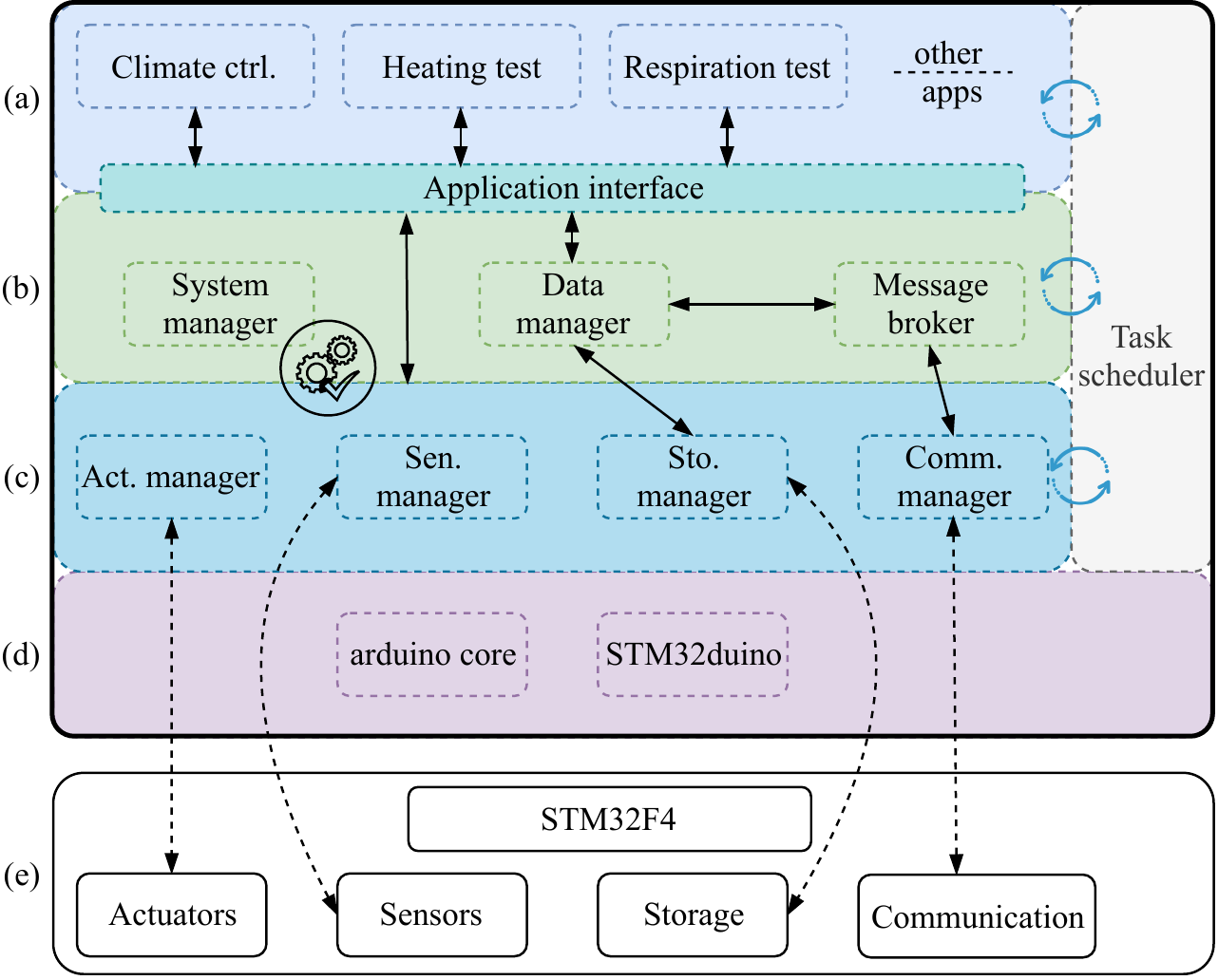}}
	\caption[firmware Arch]{Firmware architecture with its layers, components and their interaction. The layers (a) to (e) corresponds to application, middleware, device manager, hardware abstraction and hardware layer respectively. }
	\label{fig:fmw_arch}
\end{figure}

The firmware core, running on the microcontroller, is categorized into four layers to provide abstraction as seen in the Fig.~\ref{fig:fmw_arch}. The hardware abstraction layer is dependent on the microcontroller and is based on the Arduino libraries (arduino core) ported for STM-32 microcontrollers (STM32duino). The device manager layer contains managers, implemented as state machines, specific to each electronic component connected (e.g. sensors, actuators). These managers are responsible for: data access from sensors, driving the actuators, monitoring for faults, scheduling the measurements at specified intervals, intercepting external communication and logging data. The middleware layer constitutes managers which serve as brokers that direct the flow of data between the modules of different layers. Data from the sensors are accessed by the modules of different layers using a data structure managed by data manager. Message broker directs the function calls and data between the external system (e.g. decision support system or user) and the corresponding manager modules. System manager checks if all other managers are running, periodically servicing the watchdog timer, re-initializing their state machines in case of error. The task scheduler implemented is preemptive priority based, with highest priority assigned to the components of device manager layer and lower priority to the layers above it. 
 The application interface translates control decisions made by the application layer into suitable actuator commands thus abstracting the application logic from the hardware. The application layer contains the experiment regimes or apps (e.g. climate control, respiration test etc.) that are custom developed. This abstraction provides the system its programmability aspect that enables users to develop these regimes and execute them without having to deal with the underlying hardware and firmware. 

\subsubsection{Application Development}
Since the experiments performed with the chamber differs, the application that runs on the control unit must be changed accordingly. This requires the knowledge of C/C++ programming which might not be feasible for all end users. To compensate this limitation, an application development framework was developed to utilize the model-driven-development environment of MATLAB Simulink to create the experiment regimes (apps). This framework provides a Simulink interface model within which the application can be designed. Based on the designed Simulink model, this framework generates the application interface of the firmware and integrates the embedded code generated from the Simulink model into firmware core. This framework also generates an application user interface that can be directly imported into the web interface of cloud platform. This is done using a developed Matlab script that generates HTML code snippets for every input signals defined in the Simulink bus objects of the model. These code snippets are combined into standard user interface element (widget) for every bus object that can be used within the proposed cloud platform. This eliminates the need to develop user interfaces for accessing process variables and thus saving time.

\subsubsection{Cloud Framework} 
Monitoring of data in real-time and performing computation based on the current measured values are crucial. It is also necessary to scale the number of devices that can be simultaneously operated. Suitability of ThingsBoard, an open-source software, as a candidate for implementing the cloud platform can be justified by its use in several research work in the internet of things (IoT) domain \citep{Ismail2018,Paolis2018,Amir2018}. This platform includes a database server for data storage and retrieval; a web server for providing web-clients through which data and settings could be accessed or changed; and finally a protocol server (e.g., MQTT, HTTP) for enabling communication with end devices.
 
As a proof of concept, ThingsBoard was set up on a virtual machine and configured to aggregate data from multiple device instances. Interaction with the connected chamber is enabled through three configured dashboards (monitor, diagnose and control) with the possibility to import additional widgets (user-interface/UI elements). Monitor dashboard displays various measured and system parameters in real-time. Diagnosis dashboard allows manual control of actuators and also provides command line interface for debugging the designed system. Control dashboard on the other hand is application specific and allows the configuration of its parameters (e.g. set-points, control parameters etc.).

\subsubsection{Data Exchange}
Data between the server and the designed control unit are of four different types: telemetry data from the control unit to server; RPC (Remote Procedure Call) requests from server to control unit; RPC response from the control unit to server; and attributes (set-points, configs, etc.)  from server to the control unit. These four types of messages are defined as three MQTT (Message Queuing Telemetry Transport) topics (i.e., telemetry, attributes and RPC) in the server. The payload of these topics are JSON (JavaScript Object Notation) string that encapsulates the key-value pair (Listing~\ref{lst:key_val}). This key-value pair for telemetry and attributes are simply the parameter that is measured and its value. Whereas for RPC, there are only two key-value pairs of which the first corresponds to the name of the method to be called and the second the parameter that needs to be passed to the method.
\begin{lstlisting}[caption={MQTT Payload formats}, label={lst:key_val},breaklines=true,language=Python] 
# Telemetry and Attributes
{"T":21.65, "CO2":650, "ventilatorState":false, ...}
# RPC
{"method":"setTemperature", "param":"25"}
# RPC Extended
{"method":"setSetpoints", "param":"{"T_Min":20,"T_Max":25,...}"}
\end{lstlisting}

The communication between the chamber and the server is realized using an external device indicated as MQTT Streamer in Fig.~\ref{fig:sys_arch}. It forwards the JSON data received from the control units over USB to the server after encapsulating the message into a MQTT broadcast messages. The RPC calls from the server is decoded into JSON string and is forwarded to the respective control unit over USB. This forms the bridge between the standalone control unit and the server.

\subsubsection{Climate Control Application}\label{sec:climate_control}
\begin{figure}[!h]
	\centering
	\resizebox{0.485\textwidth}{!}{\includegraphics{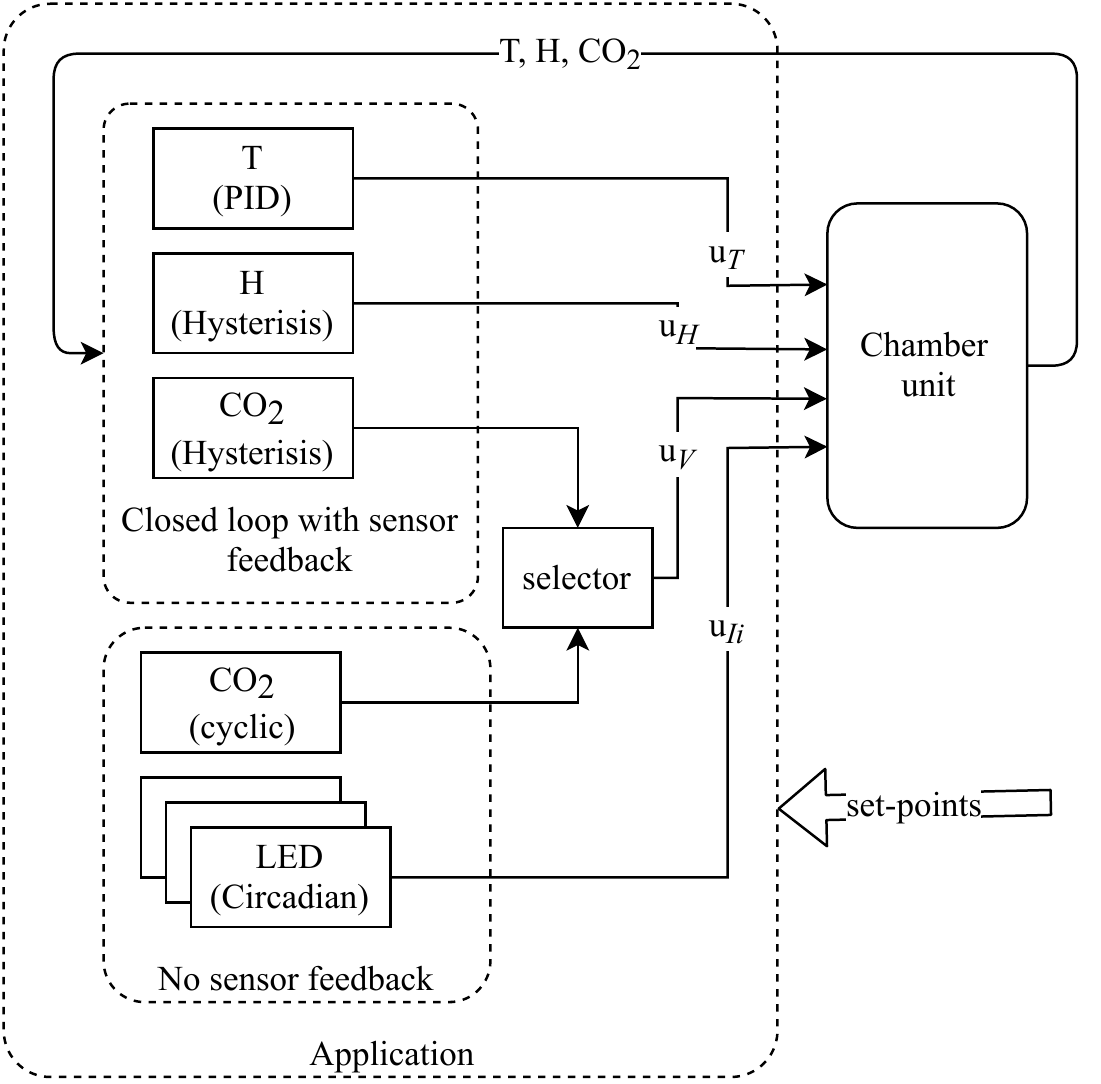}}
	\caption{Overview of the implemented example control logic for actuator control. The dashed boundary represent the climate control application integrated with the firmware core. The control signals $u_T$, $u_H$, $u_V$, and $u_{Ii}$ corresponds to the TEC, humidifier, ventilator and LEDs respectively.}
	\label{fig:controller}
\end{figure}

A set of open loop and closed loop controllers, as seen in Fig.~\ref{fig:controller}, were implemented in Simulink using the application development framework. These constitute PID controller for temperature control, hysteresis controller for humidification and CO$_2$ concentration, circadian generator for LED light spectrum and an additional cyclic controller for CO$_2$ concentration.

\paragraph{Circadian Rhythm generator}
The chamber does not include sensors for measuring the radiant flux and therefore any disturbance in this flux due to external source cannot be measured and thus not compensated. If this disturbance is avoided, the light spectrum and thus the spectral power inside the chamber can be individually varied and the necessary radiant flux can be achieved.
To achieve this, a configurable open-loop controller for circadian rhythm generation was designed. This uses a sinusoidal function to generate the base ON amplitude signal
\begin{align}
A_\mathrm{ON}(t) &= \sin\left(\pi\frac{(2t - t_\mathrm{offset})}{2t_\mathrm{on}} \right),
\end{align}
where $t_\mathrm{on}$, $t_\mathrm{off}$ and $t_\mathrm{offset}$ are the ON, OFF, and offset times. Using the given set-points for minimum and maximum percentage radiation power, the input for individual LED channels are generated as
\begin{align}
u_{Ii} =
\begin{cases}
I_{\mathrm{min}i}+I_{\mathrm{max}i} A_\mathrm{ON}(t)  & t\in[\frac{t_\mathrm{off}}{2},\frac{2t_\mathrm{on}+t_\mathrm{off}}{2}] \\
I_{\mathrm{min}i} 	&  \text{otherwise}\\
\end{cases},
\end{align}
where $I_{\mathrm{min}i}$ and $I_{\mathrm{max}i}$ are the minimum and maximum spectral power of the $i$th LED color channel and $i = 1,2,\cdots6$ corresponds to white, red, green, blue, far-red, and far-blue respectively.

\section{Results and Discussion}
Validation of the designed system followed a series of automated test routines and growth experiments with plants and insect larvae executed in the designed growth chamber. This section consolidates the results obtained from these test and provides an overview of the system capabilities and limitations.



\subsection{Temperature Sensors and TEC Performance}
\begin{figure*}[!h]
	\centering
	\includegraphics[trim=0 0 0 0,clip,scale=0.68]{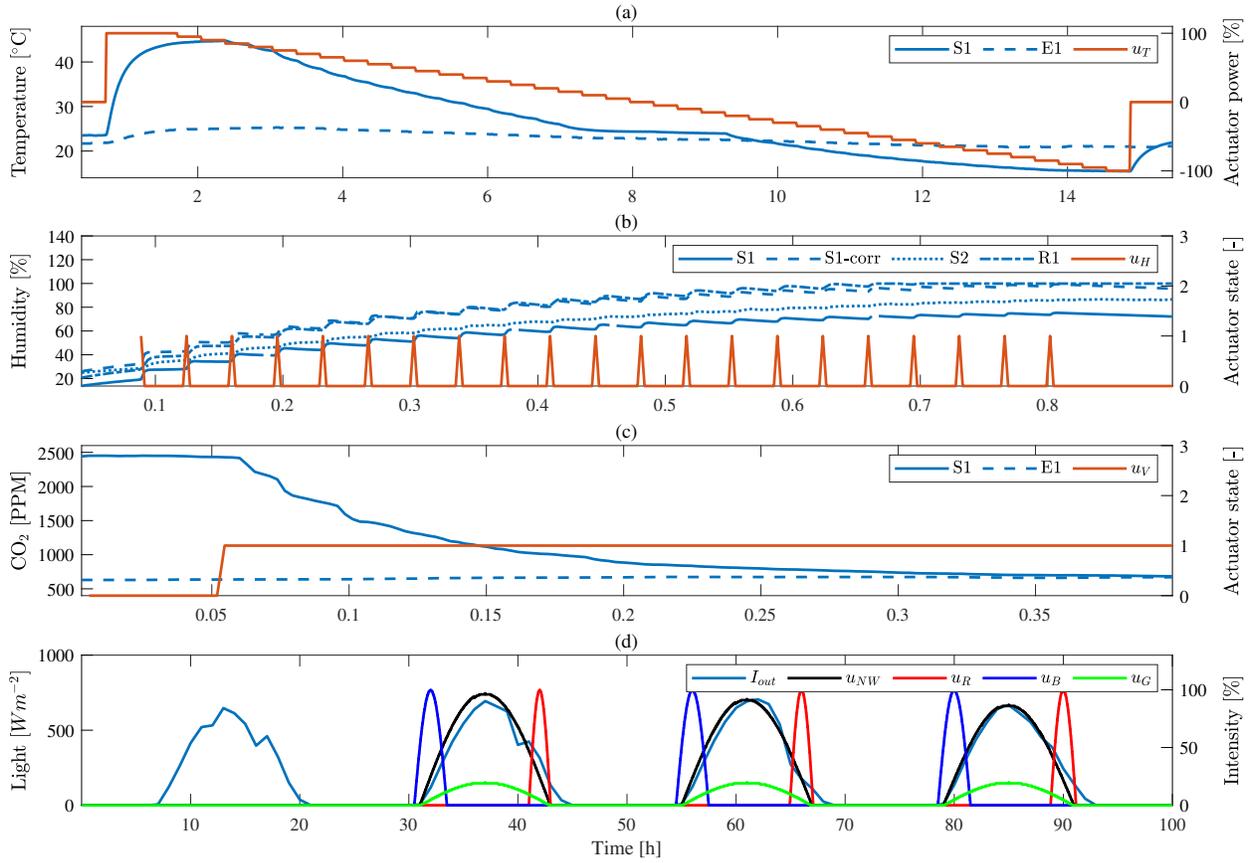}
	\caption{Evaluation of sensor and actuator performance. (a) TEC actuator and temperature sensor test. (b)Humidifier and Humidity sensor test. (c) Gas exchange test. (d) Light spectrum and circadian pattern generation. E1 and R1 corresponds to measurements made outside and inside the chamber respectively using a reference sensor. Data represented by $u$ corresponds to actuator state.}
	\label{fig:actuator_test}
\end{figure*}

An automated TEC test procedure implemented as part of the firmware was executed. This test procedure first heats the system with maximum power for a duration of \SI{1}{\hour} and then reduces the power $u_T$ in steps of 5 percent after fixed intervals. This test covers the range of operation of the TEC module in both heating and cooling modes (100 to -100\%). Temperature changes inside and outside the chamber and the corresponding power applied are recorded as seen in Fig.~\ref{fig:actuator_test}(a).

In the first hour, the temperature inside chamber rises to about \SI{44}{\celsius} and slowly saturates. When the power is reduced from 100 to 85\%, the temperature raises by about \SI{0.7}{\celsius}. This can be explained by the additional heating in the TEC due to the ripple voltage generated from the digital to analog conversion. At 100 \% power there are no voltage ripples and therefore the thermal output is reduced. The range from 20 to -20\%  power, there is no change in temperature. This behavior is again the result of digital to analog conversion where the PWM corresponding to this range does not produce voltage high enough to conduct current across the TEC. In the heating mode, the temperature inside the chamber could be raised to \si{20\celsius} above the ambient temperature. However, in cooling mode the temperature could only be lowered by \si{10\celsius}. This phenomenon could be explained by \eqref{eq:Q_in}, representing a TEC model \citep{Rowe1995,VIAN2002407}. The heat transferred by the TEC is given by
\begin{equation}\label{eq:Q_in}
Q_\mathrm{TEC}=\alpha_\mathrm{q}I_\mathrm{q}\delta T + ({I_\mathrm{q}}^2R_\mathrm{q})/2,
\end{equation}
where $\alpha_\mathrm{q}$ is the Seebeck coefficient, $I_\mathrm{q}$ is the current flowing through, $\delta T$ is the temperature difference between the hot and cold sides and $R_\mathrm{q}$ is the series resistance of the TEC module. It can be seen that the joule heating term, ${I_\mathrm{q}}^2R_\mathrm{q}$, contributes positively to the heating mode while affecting the performance negatively in the cooling mode. It can also be noticed that the outside temperature also increases during the initial heating phase. This results from the heat transferring from the chamber to the outside. It can be concluded from this observation that the thermal conductivity of the chamber walls needs to be reduced to increase the thermal insulation of the chamber.

\subsection{Humidity Sensors and Humidification}
An automated test procedure for characterizing the humidifier and testing the sensor accuracy was implemented and executed. This procedure repeatedly activates the humidifier for a fixed short duration followed by a measurement phase, where the change in humidity is monitored. Fig.~\ref{fig:actuator_test}(b) shows the course of this automated procedure and the captured measurements.

The humidity values from the sensor S1 deviates significantly from the reference measurement device (R1) due to self heating of the sensor ($\SI{3}{\celsius}$). To obtain correct readings from the sensor, it is necessary to adapt the raw data $H_\mathrm{raw}$ from the sensor as
\begin{equation}\label{eq:h_corr}
H = H_\mathrm{raw}\cdot e^{\left(17.62\cdot 243.12 \left(\frac{T_\mathrm{raw} - T}{ (243.12+T_\mathrm{raw})(243.12+T)}\right)\right)},
\end{equation}
where $T_\mathrm{raw}$ is the temperature reading from the sensor and $T = T_\mathrm{raw} -3$ is the compensated temperature. 
Data S1-Corr in Fig.~\ref{fig:actuator_test}(b), is the sensor data after applying this manufacturer recommended correction for error compensation. Sensor data S2 also shows deviation and this is due to its slow response and narrow measurement range (10-90\%RH).
The humidifier (ultrasonic atomizer) could raise the humidity inside the chamber to 100\% within 2 minutes. This translated to a conversion rate of $\SI{0.01e-3}{\kilogram\second^{-1}}$. 

\subsection{Gas Exchange}
A substrate block colonized with \textit{Lentinula edodes} fungi was placed in the chamber to raise the CO$_2$ concentration to about $2500PPM$. Concentration outside the chamber was measured around $600PPM$. Ventilation pumps were activated to reduce the concentration of CO$_2$ inside the chamber. After $\SI{18}{\min}$, the concentration of the air inside and outside the chamber equalized. This can be observed in Fig.~\ref{fig:actuator_test}(c).

The resulting volume exchange rate of the ventilator is $\SI{15}{\liter\min^{-1}}$. As a consequence of air exchange with external source, humidity inside the chamber also was reduced to that of the external source.

\subsection{Light Spectrum}
The circadian rhythm generator implemented in section \ref{sec:climate_control} was tested. The intensity for white LED channel, $I_{\mathrm{max}1}$, was set to linearly vary from 100\% to 80\% over a period of three days with no offset time. While $I_{\mathrm{max}2}$ and $I_{\mathrm{max}3}$ was fixed at 100\% for red and blue LED channel with +ve and -ve offsets respectively. The intensity for green LED channel $I_{\mathrm{max}4}$ was fixed at 20\% without offset. These amplitudes and offsets were selected such that the morning light was enriched with blue and evening light was enriched with red wavelength components. The resulting control signals $u_{NW}$, $u_{R}$, $u_{G}$ and $u_{B}$ corresponding to neutral white, red, green, and blue channels respectively is compared against the outside solar irradiation ($I_{out}$) in Fig.~\ref{fig:actuator_test}(d).

The controller was activated after day one indicating the use of offsets to adjust the day and night times within the chamber. The shape and time periods of the control signal generated for $u_{NW}$ could reproduce that of the external light. Spectrum of the light incident could be enhanced with specific wavelengths temporally using the control logic and hardware configuration. Lack of sensor feedback in the current setup limits control of the generated spectrum. Thus, any drift in spectral composition due to heating and degradation cannot be compensated.

\subsection{Plant growth}
\begin{figure*}[h]
	\centering
	\includegraphics[trim=0 0 0 0,clip,scale=0.68]{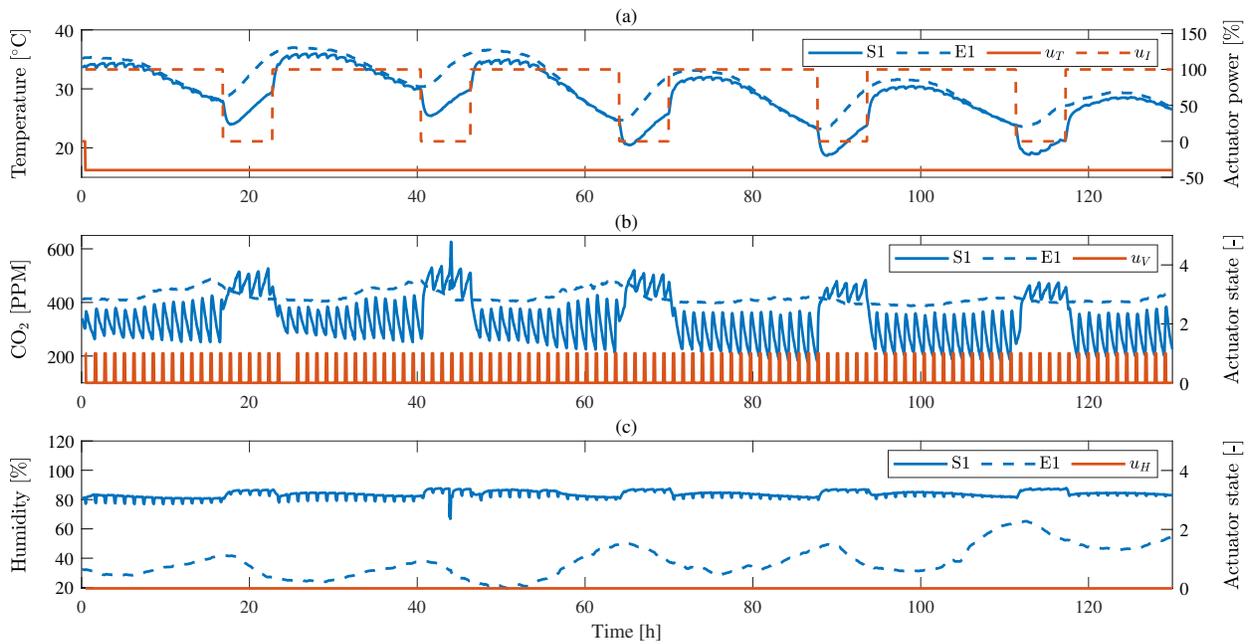}
	\caption{Plant growth experiment. (a) Temperature and Light intensity variation. (b) CO$_2$ concentration variation due to ventilation, photosynthesis and respiration. (c) Humidity variation inside and outside the chamber unit. E1 corresponds to measurements made outside the chamber using a reference sensor and data represented by $u$ corresponds to actuator state.}
	\label{fig:plant_growth}
\end{figure*}
A hydroponic growing tray consisting of about 16 \textit{Ocimum basilicum} plants of approximately 1 month old was placed in the chamber unit. Open loop controllers introduced in section \ref{sec:climate_control} was used for light gas and humidity regulation. The circadian generator was set to generate a day-night (square wave) light cycles corresponding to $\SI{18}{\hour}$ ON and $\SI{6}{\hour}$ OFF cycle. Ventilator was operated at a cyclic ON and OFF time of $\SI{120}{\sec}$ and $\SI{3420}{\sec}$ respectively. The TEC was operated at a constant power of 40\% in cooling mode to condense the transpired water. The data collected from this experiment is presented in Fig.~\ref{fig:larvae_growth} for a period of $\SI{150}{\hour}$.

The LED lighting status, indicated by $u_I$ in Fig.~\ref{fig:plant_growth}(a) shows the ON and OFF cycles. The temperature inside the chamber follows the trend of the outside temperature when the LED is ON and falls by $\SI{5}{\celsius}$ when the LED is OFF. This can be explained by the heat generated from the LED panels and the TEC operating at constant power. Variation of carbon dioxide inside and outside the chamber can be seen in Fig.~\ref{fig:plant_growth}(b). The CO$_2$ variation during the LED ON cylce, LED OFF cycle and over the entire time frame have different trend. During the LED ON cycle, the concentration drops when the ventilator is OFF and it increases when ventilator is ON. During the LED OFF cycle, the concentration increases when the ventilator is OFF and it decreases ,reaching the same concentrations as outside, when ventilator is ON. The rate change during LED ON cycle has a correlation to the temperature. These observations can be used to draw conclusions about photosynthesis and respiration taking place in these plants. It can be stated that light and temperature influence plant photosynthesis resulting in production of CO$_2$. Net CO$_2$ produced in the absence of light is negative which indicates the respiration process. The plant species under study shows increase in photosynthesis rate with decrease in temperatures. 
Humidity, as seen in Fig.~\ref{fig:plant_growth}(c), stays constant during the LED ON cycle and is slightly elevated during the OFF cycle indicating higher transpiration.

\subsection{Larvae growth}
\begin{figure*}[!h]
	\centering
	\includegraphics[trim=0 0 0 0,clip,scale=0.68]{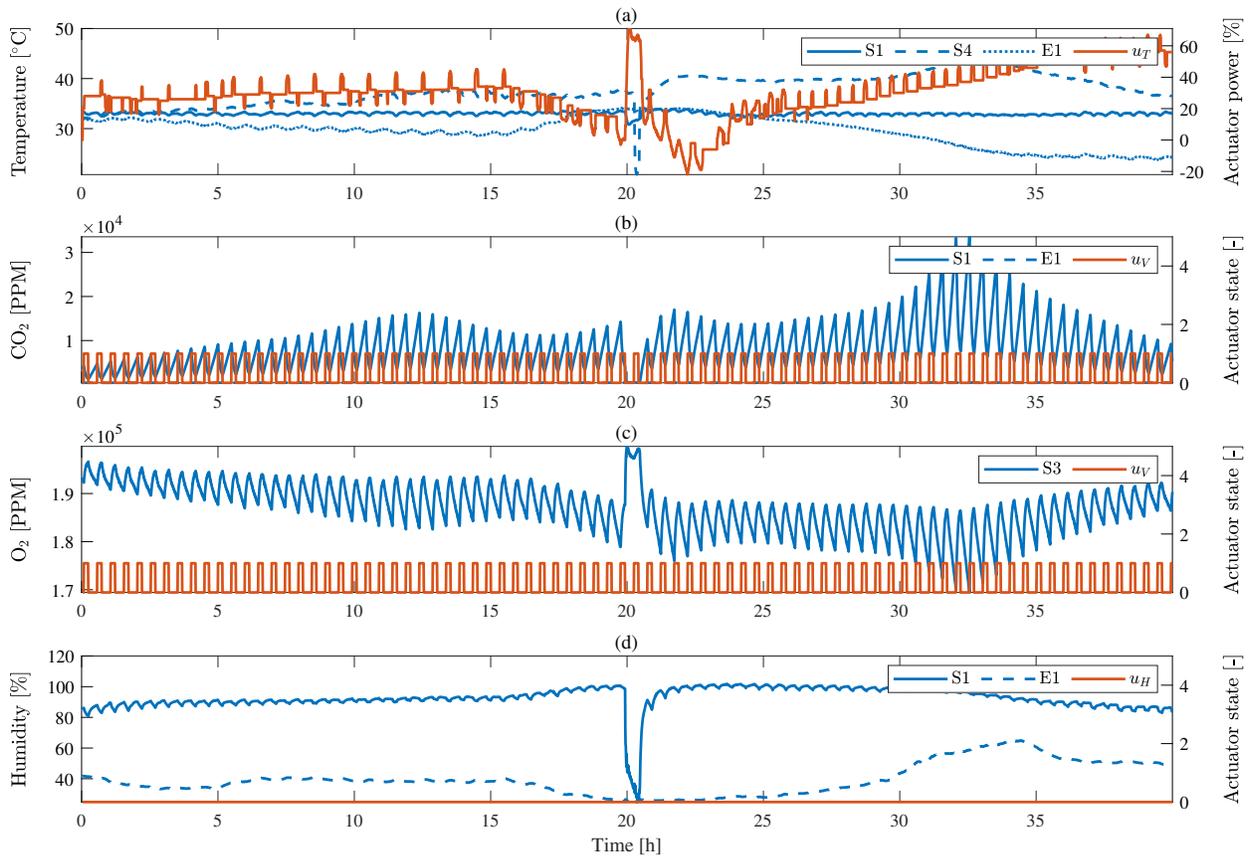}
	\caption{Larvae growth experiment. (a) Temperature variation in growing medium, chamber unit and outside. (b and c) CO$_2$ and O$_2$ concentration variation due to ventilation and larval metabolic activity. (d) Humidity variation inside and outside the chamber unit. E1 corresponds to measurements made outside the chamber using a reference sensor and data represented by $u$ corresponds to actuator state.}
	\label{fig:larvae_growth}
\end{figure*}

About 2000 hatchlings of \textit{Hermetia illucens} were introduced in a growing tray containing a substrate made of chicken feed and water mixture. The growing tray was placed in the chamber unit with sensor S4 and S5 submerged in the substrate. The example climate control application introduced in section \ref{sec:climate_control} was used as the application running on the control unit. A temperature set point of $\SI{33}{\celsius}$ was used to regulate the air temperature inside the chamber. Cyclic controller for ventilator was selected with an ON and OFF time of $\SI{600}{\sec}$ and $\SI{1200}{\sec}$ respectively to regulate the CO$_2$ concentration.
The data collected using the designed system is presented in Fig.~\ref{fig:larvae_growth} for a period of $\SI{40}{\hour}$.

The air temperature inside chamber is maintained at the set $\SI{33}{\celsius}$ as seen in Fig.~\ref{fig:larvae_growth}(a). The PID controller compensating the variation in temperature due to the ventilation can be seen in the peaks of the control signal $u_T$. It can be noticed that the substrate temperature (S4) gradually increases and reaches a peak temperature of $\SI{45.5}{\celsius}$ at $\SI{35}{\hour}$. Similarly, the variation in the CO$_2$ and O$_2$ concentrations increases gradually also reaching the peak around the same time step. These variations in substrate temperature and gas concentrations can be concluded to be results of larval metabolism. Correlating these three measurements, it can be stated that increase in temperature increases larval metabolic activity and thus the dependency of the larval growth on temperatures can be concluded. Another observation that can be made from the humidity measurement (Fig.~\ref{fig:larvae_growth}(d)) of air is that, after the time step corresponding to the peak metabolic activity, humidity gradually falls. Moisture in substrate is gradually lost through evaporation and consumption by the larvae. It can be stated that decrease in air humidity can be explained by the substrate drying up. It is however inconclusive from this experiment if the fall in larval metabolic activity is due to the dry substrate.

\section{Conclusion}

Design of a programmable growth chamber using off-the-shelf components, custom designed 3D printed parts, open-source and self developed software was possible.
Actuators for resource exchange, climate control and light spectrum adjustment were designed, incorporated and tested. 
Software framework was developed and tested for programming experiments and logging sensor data and system states.
This framework also enabled accessing logged data, adjusting controller parameters and performing diagnostics through a graphical interface both locally and remotely. 
Example applications were developed using the application development framework to showcase the capability of the system to execute custom control algorithms and experiment procedures for automated study and information capture.
Results of the actuator and sensor tests revealed their capabilities, limitation and also improvement areas such as chamber insulation for better thermal performance.
These results also indicated that sensors S1-S4 performed well in measuring the required parameters while S5 could be replaced for a more robust moisture sensor when available.
Experiments performed with plants and larvae generated data that can be used to perform quantitative studies on the biomass and byproduct production.
These experiments also provided data that helps study the growth processes and its responses to the environmental factors.


The designed system fulfills the requirements of a growth chamber that is programmable, low-cost ($<1000$ EUR) and open-source based suitable for small plants, insect larvae and mushrooms. The results obtained and presented in this work serves as fundamental setup for the detailed study of individual subjects and development of mathematical models which shall be the focus of future work.
\bibliographystyle{elsarticle-num} 
\bibliography{./bib/main_bib}

\begin{thebibliography}{10}
\expandafter\ifx\csname url\endcsname\relax
  \def\url#1{\texttt{#1}}\fi
\expandafter\ifx\csname urlprefix\endcsname\relax\def\urlprefix{URL }\fi
\expandafter\ifx\csname href\endcsname\relax
  \def\href#1#2{#2} \def\path#1{#1}\fi

\bibitem{DESPOMMIER2011}
D.~D. Despommier, The vertical farm : feeding the world in the 21st century,
  St. Martin’s Press, New York, 2011.
\newblock \href {https://doi.org/B-978-0-312-61139-2}
  {\path{doi:B-978-0-312-61139-2}}.

\bibitem{KOZAI2013}
T.~Kozai, Sustainable plant factory: Closed plant production systems with
  artificial light for high resource use efficiencies and quality produce, Acta
  Horticulturae 1004 (2013) 27--40.
\newblock \href {https://doi.org/10.17660/ActaHortic.2013.1004.2}
  {\path{doi:10.17660/ActaHortic.2013.1004.2}}.

\bibitem{ALCHALABI201574}
M.~{Al-Chalabi}, Vertical farming: Skyscraper sustainability?, Sustainable
  Cities and Society 18 (2015) 74 -- 77.
\newblock \href {https://doi.org/https://doi.org/10.1016/j.scs.2015.06.003}
  {\path{doi:https://doi.org/10.1016/j.scs.2015.06.003}}.

\bibitem{LASSEUR2005}
C.~Lasseur, C.~Paillé, B.~Lamaze, P.~Rebeyre, A.~Rodriguez, L.~Ordonez,
  F.~Marty, Melissa: Overview of the project and perspectives, in: SAE
  Technical Paper, SAE International, 2005.
\newblock \href {https://doi.org/10.4271/2005-01-3066}
  {\path{doi:10.4271/2005-01-3066}}.

\bibitem{BOCKSTAHLER2017}
K.~Bockstahler, R.~Hartwich, C.~Matthias, J.~Witt, S.~Hovland, D.~Laurini,
  Status of the advanced closed loop system acls for accommodation on the iss,
  International Conference on Environmental Systems 135.

\bibitem{CARY1994}
C.~Mitchell, Bioregenerative life-support systems, The American Journal of
  Clinical nutrition 60 (1994) 820S--824S.
\newblock \href {https://doi.org/10.1093/ajcn/60.5.820S}
  {\path{doi:10.1093/ajcn/60.5.820S}}.

\bibitem{LOBASCIO2008}
C.~Lobascio, M.~Lamantea, V.~Cotronei, B.~Negri, S.~De~Pascale, A.~Maggio,
  M.~Maffei, M.~Foti, S.~Palumberi, Cab: The bioregenerative life support
  system. a feasibility study on the survivability of humans in a long-duration
  space missions, Acta Horticulturae 801 (2008) 465--472.
\newblock \href {https://doi.org/10.17660/ActaHortic.2008.801.50}
  {\path{doi:10.17660/ActaHortic.2008.801.50}}.

\bibitem{KITAYA1994}
Y.~Kitaya, A.~Tani, M.~Kiyota, I.~Aiga, Plant growth and gas balance in a plant
  and mushroom cultivation system, Advances in Space Research 14~(11) (1994)
  281 -- 284.
\newblock \href {https://doi.org/https://doi.org/10.1016/0273-1177(94)90309-3}
  {\path{doi:https://doi.org/10.1016/0273-1177(94)90309-3}}.

\bibitem{Jagath2010}
J.~W. Rupasinghe, J.~O.~S. Kennedy, Economic benefits of integrating a
  hydroponic-lettuce system into a barramundi fish production system,
  Aquaculture Economics \& Management 14~(2) (2010) 81--96.
\newblock \href {https://doi.org/10.1080/13657301003776631}
  {\path{doi:10.1080/13657301003776631}}.

\bibitem{CONRAD2017}
Z.~Conrad, S.~Daniel, V.~Vincent, Vertical farm 2.0: Designing an economically
  feasible vertical farm - a combined european endeavor for sustainable urban
  agriculture, Tech. rep., Association for Vertical Farming (2017).

\bibitem{CUBESCircle2018}
C.~Circle, \href{https://www.cubescircle.de/}{Cubes cirlce: Future food
  production} (2018).
\newline\urlprefix\url{https://www.cubescircle.de/}

\bibitem{CONVIRON2019}
Conviron, GEN1000 REACH-IN multi-application chamber, Conviron, rev 04 (2019).

\bibitem{PFC2019}
E.~Castell{\'o}~Ferrer, J.~Rye, G.~Brander, T.~Savas, D.~Chambers, H.~England,
  C.~Harper, Personal food computer: A new device for controlled-environment
  agriculture, in: K.~Arai, R.~Bhatia, S.~Kapoor (Eds.), Proceedings of the
  Future Technologies Conference (FTC) 2018, Springer International Publishing,
  Cham, 2019, pp. 1077--1096.

\bibitem{ZABEL20161}
P.~Zabel, M.~Bamsey, D.~Schubert, M.~Tajmar, Review and analysis of over 40
  years of space plant growth systems, Life Sciences in Space Research 10
  (2016) 1 -- 16.
\newblock \href {https://doi.org/https://doi.org/10.1016/j.lssr.2016.06.004}
  {\path{doi:https://doi.org/10.1016/j.lssr.2016.06.004}}.

\bibitem{Philippoussis2003}
A.~Philippoussis, P.~Diamantopoulou, G.~Zervakis, Correlation of the properties
  of several lignocellulosic substrates to the crop performance of the shiitake
  mushroom {Lentinula edodes}, World Journal of Microbiology and Biotechnology
  19~(6) (2003) 551--557.
\newblock \href {https://doi.org/10.1023/A:1025100731410}
  {\path{doi:10.1023/A:1025100731410}}.

\bibitem{John1990}
J.~E. Erwin, R.~D. Heins, Temperature effects on lily development rate and
  morphology from the visible bud stage until anthesis, Journal of the American
  Society for Horticultural Science jashs 115~(4) (1990) 644 -- 646.

\bibitem{BLOCK1953}
S.~S. Block, Humidity requirements for mold growth, Applied microbiology 1~(6)
  (1953) 287--293.

\bibitem{KULWICKI1984}
B.~Kulwicki, Ceramic sensors and transducers, Journal of Physics and Chemistry
  of Solids 45~(10) (1984) 1015 -- 1031.
\newblock \href {https://doi.org/https://doi.org/10.1016/0022-3697(84)90046-5}
  {\path{doi:https://doi.org/10.1016/0022-3697(84)90046-5}}.

\bibitem{Paganelli1984}
J.~Paganelli, Ptc ceramic heaters in automotive controls, in: SAE Technical
  Paper, SAE International, 1984.
\newblock \href {https://doi.org/10.4271/840143} {\path{doi:10.4271/840143}}.

\bibitem{CHEIN20042207}
R.~Chein, G.~Huang, Thermoelectric cooler application in electronic cooling,
  Applied Thermal Engineering 24~(14) (2004) 2207 -- 2217.
\newblock \href
  {https://doi.org/https://doi.org/10.1016/j.applthermaleng.2004.03.001}
  {\path{doi:https://doi.org/10.1016/j.applthermaleng.2004.03.001}}.

\bibitem{RIFFAT20041979}
S.~Riffat, G.~Qiu, Comparative investigation of thermoelectric air-conditioners
  versus vapour compression and absorption air-conditioners, Applied Thermal
  Engineering 24~(14) (2004) 1979 -- 1993.
\newblock \href
  {https://doi.org/https://doi.org/10.1016/j.applthermaleng.2004.02.010}
  {\path{doi:https://doi.org/10.1016/j.applthermaleng.2004.02.010}}.

\bibitem{TOPP1972127}
M.~Topp, P.~Eisenklam, Industrial and medical uses of ultrasonic atomizers,
  Ultrasonics 10~(3) (1972) 127 -- 133.
\newblock \href {https://doi.org/https://doi.org/10.1016/0041-624X(72)90009-1}
  {\path{doi:https://doi.org/10.1016/0041-624X(72)90009-1}}.

\bibitem{Andria2016}
G.~{Andria}, G.~{Cavone}, C.~G.~C. {Carducci}, M.~{Spadavecchia}, A.~{Trotta},
  A pwm temperature controller for themoelectric generator characterization,
  in: 2016 IEEE Metrology for Aerospace (MetroAeroSpace), 2016, pp. 291--296.
\newblock \href {https://doi.org/10.1109/MetroAeroSpace.2016.7573229}
  {\path{doi:10.1109/MetroAeroSpace.2016.7573229}}.

\bibitem{OGUNTOYINBO2013340}
B.~Oguntoyinbo, T.~Ozawa, K.~Kawabata, J.~Hirama, H.~Yanagibashi, Y.~Matsui,
  A.~Kurahashi, T.~Shimoda, M.~Taniguchi, K.~Nishibori, Sma (speaking mushroom
  approach) environmental control system development: Automated cultivation
  control system characterization, IFAC Proceedings Volumes 46~(4) (2013) 340
  -- 345.
\newblock \href
  {https://doi.org/https://doi.org/10.3182/20130327-3-JP-3017.00077}
  {\path{doi:https://doi.org/10.3182/20130327-3-JP-3017.00077}}.

\bibitem{Schwend2016243}
T.~Schwend, D.~Prucker, S.~Peisl, A.~Nitsopoulos, H.~Mempel, The rosmarinic
  acid content of basil and borage correlates with the ratio of red and far-red
  light, European Journal of Horticultural Science 81 (2016) 243--247.
\newblock \href {https://doi.org/DOI:
  https://doi.org/10.17660/eJHS.2016/81.5.2} {\path{doi:DOI:
  https://doi.org/10.17660/eJHS.2016/81.5.2}}.

\bibitem{KIM2004143}
S.-J. Kim, E.-J. Hahn, J.-W. Heo, K.-Y. Paek, Effects of leds on net
  photosynthetic rate, growth and leaf stomata of chrysanthemum plantlets in
  vitro, Scientia Horticulturae 101~(1) (2004) 143 -- 151.
\newblock \href {https://doi.org/https://doi.org/10.1016/j.scienta.2003.10.003}
  {\path{doi:https://doi.org/10.1016/j.scienta.2003.10.003}}.

\bibitem{Schwend2016148}
T.~Schwend, M.~Kriedel, D.~Prucker, S.~Peisl, H.~Mempel, On the role of the
  light regime in root development of euphorbia pulcherrima leafy stem
  cuttings, European Journal of Horticultural Science 81 (2016) 148--151.
\newblock \href {https://doi.org/DOI:
  https://doi.org/10.17660/eJHS.2016/81.3.2} {\path{doi:DOI:
  https://doi.org/10.17660/eJHS.2016/81.3.2}}.

\bibitem{Massa01122008}
G.~D. Massa, H.-H. Kim, R.~M. Wheeler, C.~A. Mitchell, Plant productivity in
  response to led lighting, HortScience 43~(7) (2008) 1951--1956.

\bibitem{XPELED2018}
Cree, Cree® XLamp® XP-E LEDs, Cree, rev. 25A (2018).

\bibitem{MCELED2018}
Cree, Cree® XLamp® MC-E LED, Cree, rev. 11N (2018).

\bibitem{Ismail2018}
A.~A. {Ismail}, H.~S. {Hamza}, A.~M. {Kotb}, Performance evaluation of open
  source iot platforms, in: 2018 IEEE Global Conference on Internet of Things
  (GCIoT), 2018, pp. 1--5.
\newblock \href {https://doi.org/10.1109/GCIoT.2018.8620130}
  {\path{doi:10.1109/GCIoT.2018.8620130}}.

\bibitem{Paolis2018}
L.~T. {De Paolis}, V.~{De Luca}, R.~{Paiano}, Sensor data collection and
  analytics with thingsboard and spark streaming, in: 2018 IEEE Workshop on
  Environmental, Energy, and Structural Monitoring Systems (EESMS), 2018, pp.
  1--6.
\newblock \href {https://doi.org/10.1109/EESMS.2018.8405822}
  {\path{doi:10.1109/EESMS.2018.8405822}}.

\bibitem{Amir2018}
S.~{Amir Alavi}, A.~{Rahimian}, K.~{Mehran}, J.~{Mehr Ardestani}, An iot-based
  data collection platform for situational awareness-centric microgrids, in:
  2018 IEEE Canadian Conference on Electrical Computer Engineering (CCECE),
  2018, pp. 1--4.
\newblock \href {https://doi.org/10.1109/CCECE.2018.8447718}
  {\path{doi:10.1109/CCECE.2018.8447718}}.

\bibitem{Rowe1995}
D.~M. Rowe, CRC handbook of thermoelectrics, CRC Press, Boca Raton, FL, 1995.

\bibitem{VIAN2002407}
J.~Vi\'an, D.~Astrain, M.~Dom\'inguez, Numerical modelling and a design of a
  thermoelectric dehumidifier, Applied Thermal Engineering 22~(4) (2002) 407 --
  422.
\newblock \href {https://doi.org/https://doi.org/10.1016/S1359-4311(01)00102-8}
  {\path{doi:https://doi.org/10.1016/S1359-4311(01)00102-8}}.

\end{thebibliography}
\end{document}